\newcommand{\EQ}{\begin{equation}}
\newcommand{\EE}{\end{equation}}

\newcommand{\s}{{\rm stat}}
\newcommand{\siml}{\,\raisebox{-.6ex}{$\stackrel{<}{\scriptstyle{\sim}}$}\,}
\newcommand{\simg}{\,\raisebox{-.6ex}{$\stackrel{>}{\scriptstyle{\sim}}$}\,}
\newcommand{\ep}{\varepsilon}
\renewcommand{\a}{\alpha}
\renewcommand{\max}{{\rm m}}

\documentclass[10pt]{bmc_article}

\usepackage{cite} 
\usepackage{url}  
\usepackage{ifthen}  
\usepackage{multicol}   
\usepackage{epsfig}
\usepackage[applemac]{inputenc} 
\newboolean{publ}

\newenvironment{bmcformat}{\fussy\setboolean{publ}{true}}{\fussy}

\begin{document}
\begin{bmcformat}

\title{Adaptive evolution of transcription factor binding sites}
 
\author{
        Johannes Berg$^{1}$%
        \email{Johannes Berg - berg@thp.uni-koeln.de}%
      \and
        Stana Willmann$^1$%
        \email{Stana Willmann - sr@thp.uni-koeln.de}%
        and
        Michael L\"assig\correspondingauthor$^1$%
        \email{Michael L\"assig\correspondingauthor - lassig@thp.uni-koeln.de}
       }

\address{%
    \iid(1)Institute for Theoretical Physics,\\
    University of Cologne, \\
    50937 Cologne,\\
    Germany
  }%

\maketitle

\noindent
{\it BMC Evolutionary Biology} {\bf 4}(1):42 (2004)

\begin{abstract}
        \paragraph*{Background:} 
The regulation of a gene depends on the
binding of transcription factors to specific sites located
in the regulatory region of the gene. The
generation of these binding sites and of cooperativity 
between them are essential building blocks in the
evolution of complex regulatory networks. We study a
theoretical  model for the sequence evolution of
binding sites by point mutations. The approach is based 
on biophysical models for the binding of transcription 
factors to DNA. Hence we derive empirically grounded 
fitness landscapes, which enter a population genetics 
model including mutations, genetic drift, and selection. 
\paragraph*{Results:} 
We show that the selection for factor binding generically
leads to specific correlations between nucleotide frequencies
at different positions of a binding site.  
We demonstrate the possibility of rapid adaptive 
evolution generating a new binding site for
a given transcription factor by point mutations. 
The evolutionary time required is
estimated  in terms of the neutral (background) mutation rate, the
selection coefficient, and the effective population size.
\paragraph*{Conclusions:} 
The efficiency of binding site formation is seen to depend on two 
joint conditions:  
the binding site motif must be short enough and the promoter region must 
be long enough. These constraints on promoter architecture are indeed seen in 
eukaryotic systems. Furthermore, we analyse the adaptive evolution of
genetic switches and of signal integration through  binding
cooperativity  between different sites. Experimental tests
of this picture involving the statistics of polymorphisms
and phylogenies of sites are discussed.
\end{abstract}

\ifthenelse{\boolean{publ}}{\begin{multicols}{2}}{}

\section*{Background}
The expression of a gene is controlled by other genes expressed at the
same time and by external signals, a process called {\em gene 
regulation}~\cite{PtashneGann.book:2002}.  Due to the combinatorial
complexity of regulation, a large number of functional tasks can be
performed by a limited number of genes. Differences in gene regulation
are believed to be a major source of diversity in higher eukaryotes.

To a large extent, gene regulation is the control of
transcription.  It is accomplished by a number of
regulatory proteins called  {\em transcription factors}
that bind to specific sites on DNA. These binding sites
contain about $10-15$ base
pairs relevant for binding and are mostly located in the
cis-regulatory promoter region of a gene. A
cis-regulatory region  in {\it E.~coli} is about $300$ base
pairs long and contains a few  transcription factor
binding  sites~\cite{Collado_VidesMagasanik:1991}. There may
be two or more sites for the same factor in one
promoter region.  At the same time, the sequences of
binding sites are {\em fuzzy}, that is, different sites for
the same factor differ by about $20-30$ percent of the
bases relevant for binding~\cite{Collado_VidesMagasanik:1991}. 
This makes the identification of 
sites a difficult bioinformatics
problem~\cite{BussemakerSiggia:2000,HertzStormo:1999,StormoFields:1998}.
Frequently,  the simultaneous binding at two nearby sites
is energetically favoured. This  so-called {\em binding
cooperativity} can be related to  various functions. In a 
{\em genetic switch} such as the famous phage lambda switch in
{\em Escherichia~coli}~\cite{Ptashne.book:1992}, it produces a sharp
increase  of the expression level at a certain threshold
concentration of a transcription factor. A pair of  sites
for two different kinds of factors with  cooperative
binding can be  a simple module for {\em signal
integration}, leading to the expression of the downstream
gene only when both kinds of factors are present
simultaneously~\cite{PtashneGann.book:2002}. These examples are
discussed in more detail below. Regulation in higher
eukaryotes shares these features but is vastly more
complicated~\cite{StoneWray:2001}. A promoter region is typically a few
thousand base pairs long and contains many  different
binding sites with often complex interactions. At the same time,
individual sites are shorter, with about 5-8 relevant base pairs. The
sites are sometimes organized in {\em modules} interspersed between 
regions containing no sites. In many
known cases, the expression of a gene depends on the
simultaneous presence of several factors. Well-studied
examples of regulatory networks in eukaryotes include the
sea urchin  {\it Strongylocentrotus purpuratussea}
\cite{Davidson:1999} and the early developmental genes  in {\it
Drosophila} \cite{Tautz:2000}.

The sequence statistics of binding sites has been addressed
in two recent theoretical
studies~\cite{GerlandHwa:2002,SenguptaDjordjevicShraiman:2002}. 
Based on a model incorporating the biophysics of sequence-factor
interaction~\cite{BergvonHippel:1987,GerlandMorozHwa:2002}, a 
{\em fitness landscape} for binding site sequences is
constructed (see the discussion in the next section). 
The resulting mutation-selection equilibrium is
analysed using a mean-field {\em quasispecies}
approach~\cite{EigenMcCaskillSchuster:1989}. This approach,
which neglects the effects of genetic drift, is applicable in
very large populations. In both
studies~\cite{GerlandHwa:2002,SenguptaDjordjevicShraiman:2002}, fuzziness is
attributed to {\em mutational entropy} as a possible reason: the single or 
few sequence states with optimal binding of the transcription factor can
be outweighed by the vastly higher number of sub-optimal states 
at some mutational distance from the optimal binding sequence. This effect is 
similar to the fuzziness of amino acid
sequences in proteins discussed in~\cite{GoldsteinLuthey-SchultenWolynes:1992}.

~From an evolutionary perspective, explaining the molecular
programming of regulatory networks presents a striking problem.
The diversification of higher eukaryotes, in
particular, requires the efficient generation and
alteration of  regulatory binding interactions.
One likely mode of evolution is gene duplications
with subsequent complementary {\em losses of function}
in both copies~\cite{Wagner:2002,LynchO'HelyWalshForce:2001}. However, the
differentiation of regulation should also require
complementary 
processes that generate {\em new functions} of genes
as a response to specific  demands. This task must be
accomplished mainly by sequence
evolution of regulatory DNA.  There are examples 
of highly conserved regulatory sequences with a conserved 
function but binding sites can also 
appear, disappear, or alter their sequence even
between relatively closely related 
species; see, e.g., refs.
~\cite{LudwigKreitman:1995,LudwigPatelKreitman:1998,DermitzakisBergmanClark:2002,Scemama.etal:2002,Arnosti:2003}.
This turnover of binding sites 
 has been argued to follow an approximate molecular 
clock in {\em Drosophila}~\cite{CostasCasaresVieira:2003}.
The transcription factors
themselves are known to remain more conserved, especially
if they are involved in the regulation of more than one
gene. 

The modes of regulatory sequence evolution and their relative
importance remain largely to be explored. Contributions
may arise from point mutations, slippage processes~\cite{McGregor.etal:2001},
and larger rearrangements of promoter regions~\cite{Shapiro:1999}.
The latter processes may lead to the shuffling of entire modules
of binding sites between different genes. 
In this paper, we are more interested in the local sequence evolution
within a module, which has been argued to contribute most of the 
promoter sequence difference between species~\cite{Wray.etal:2003}. It is also
the most promising starting point for a {\em quantitative} analysis 
of binding site evolution. We study a theoretical model that takes
into account point mutations,
selection, and genetic drift. The form of selection is
inferred from the biophysics of the binding interactions
between transcription factors and DNA. 

We derive the stationary distribution of binding sites under 
selection, which shows specific correlations between nucleotide 
frequencies at different positions in a binding site. The 
non-stationary solutions of the model  describe   
efficient adaptive pathways for
the molecular evolution of regulatory networks by point 
mutations. This 
efficiency can be quantified in terms of the length 
of the binding motif, and the length of the promoter 
region, and the 
fitness landscape for factor binding, which is amenable to 
quite explicit modeling.  
 
With the parameters found in natural systems, 
our model predicts that a new
binding site for a given transcription factor can be
generated by a fast series of adaptive 
substitutions, even if the expression of the
corresponding gene bears even a modest fitness advantage. The
evolutionary time required for site formation in response to 
a {\em newly arising} selection pressure is estimated in terms of the
characteristic time scales of mutation, selection, and
drift. For  {\it Drosophila}, it may be as short as 
$10^5$ years even for moderate selection pressures. 
However, this pathway is found to depend crucially
on the presence of selection. It would be too slow under 
neutral evolution, in contrast to the results of~\cite{StoneWray:2001}, see also the 
recent discussion in \cite{MacArthurBrookfield:2004}.   
Cooperative interactions between binding
sites can evolve adaptively on similar time scales, as we show for
the two simple examples alluded to above, the genetic
switch and the signal integration module. These results
are discussed at the end of the paper with particular
emphasis on possible experimental tests.

\section*{Factor binding and selection}

The binding energy (measured in units of $k_B T$) between a transcription 
factor and its
binding site is, to a good approximation, the sum of
independent contributions from a small number of important
positions of the binding site sequence, $E/k_B T =
\sum_{i=1}^\ell \ep_i$, with $\ell \approx
10-15$~\cite{Fields.etal:1997,Oda.etal:1998,SaraiTakeda:1989}. 
The individual contributions
$\ep_i$ depend on the position $i$ and on the nucleotide
$a_i$ at that position. There is typically one particular
nucleotide $a_i^*$ preferred for binding; the sequence
$(a_1^*, \dots, a_\ell^*)$ is called the {\em target 
sequence}. The target sequence can be inferred as the
consensus sequence of a sufficiently large number of
equivalent sites. The so-called {\em energy matrix} 
$\ep_i(a)$ has been determined experimentally for some factors from
{\em in vitro} measurements of the binding affinity for
each single-nucleotide mutant of the target sequence. 
Typical values for the loss in binding energy are 1-3 $k_B
T$ per single-nucleotide mismatch away from the target
sequence. In this paper, we use the further approximation $\ep_i =
\ep$ if $a_i = a_i^*$ and $\ep = 0$ otherwise, the
so-called {\em two-state model}~\cite{BergvonHippel:1987}. The
binding energy of any sequence $(a_1, \dots, a_\ell)$ is
then, up to an irrelevant constant, simply given by its
Hamming distance $r$ to the target sequence: $E/k_B T  = \ep r$.
(The Hamming distance is defined as the number of positions
with a mismatch $a_i \neq a_i^*$.) 

It is important to note
the status of this ``minimal model'' of binding energies
for the discussion in this
paper. Both approximations underlying the model can be 
violated. Even though typical mismatch energies are of the same
order of magnitude, there can be considerable differences
between different substitutions at one position and
between different nucleotide positions. Moreover,
deviations from the approximate additivity of binding
energies for the single nucleotide positions have also been 
observed. However, these complications do not affect the
order-of-magnitude estimates for adaptive sequence
evolution. As it will become clear, the efficiency
of binding site formation depends only on the qualitative shape
of the fitness landscapes derived below. In these
landscapes, the regime of weakly-binding sequences and
of strongly-binding sequences are separated by only a few
single nucleotide substitutions. The relative magnitude
of the fitness increase of these substitutions does
not matter in first approximation. Indeed, inhomogeneities
in the values of the $\ep_i(a)$ tend to reduce the number 
of {\em crucial} steps in the adaptive process and thereby 
to further increase its speed.

Within the two-state model, the  binding
probability of the factor in thermodynamic equilibrium is
\EQ
p = \frac{1}{1 + \exp[\ep (r - \rho)]}.
\label{p}
\EE
Here $\ep$ is the binding energy per nucleotide mismatch and $\ep \rho$
is the chemical potential measuring the factor concentration.
Both parameters are expressed in units of $k_B T$ and hence dimensionless.
Appropriate values for typical binding sites have
been discussed extensively in
refs.~\cite{GerlandHwa:2002,GerlandMorozHwa:2002}. It is found 
that $\ep$ should take values around $2$,
which is consistent with the  measurements for
known transcription factors mentioned 
above~\cite{Fields.etal:1997,Oda.etal:1998,SaraiTakeda:1989}. 
The chemical potential 
depends  on the number of transcription factors present in
the cell, on the binding probability to
{\em background} sites elsewhere in the genome (which have
a sequence similar to the target sequence by chance), and
on the {\em functional} sites in the in the 
genome other than the binding site in question
that may compete for the same protein. Binding to 
background sites does not significantly reduce the binding 
to a specific functional site~\cite{GerlandMorozHwa:2002}.
This leads to  values
$ \rho \approx (\log n_f) /\ep \approx
2-4$, given observed factor numbers  
$n_f$ of about $50 - 5000$ ~\cite{GerlandMorozHwa:2002}. 
Binding to other copies
of the same functional sequence becomes only relevant at low
factor concentrations and high number of copies, when sites
compete for factors.

A {\em fitness landscape} quantifies the fitness contribution 
$F(a_1,\dots, a_\ell)$ of each sequence state at the binding site.
Fitness differences arise due to different expression levels
of the regulated gene, and these in turn depend on the binding
of the transcription factors. It is only these fitness differences 
that enter the population dynamics of binding site sequences in the next 
section. Following the conceptual
framework of ref.~\cite{GerlandHwa:2002}, we assume that the
environment of the regulated gene  can be described by
a number of {\em cellular states} (labelled by the index
$\alpha$) with  different transcription factor
concentrations, i.e., with different chemical potentials
$\rho^\a$. These cellular states can be thought of as different stages
within a cell cycle. In each state, the fitness depends on
the expression level of the regulated gene in a specific way. 
This expression level is 
determined by the binding probability $p^\a$ of the
transcription factor.  Assuming  that both dependencies are
linear (this is not crucial) and that the cellular states
contribute additively to the overall fitness $F$, we
obtain
\EQ
F = \sum_\a s^\a p^\a.
\label{F1}
\EE
Here the {\em selection coefficient} $s^\a$ is defined as 
the fitness difference (due to different expression of the downstream 
gene) between the cases of complete factor binding
and no binding in the state $\a$.
Such fitness differences can now
be measured directly in viral systems~\cite{Opijnen_etal:2004}. 
Inserting (\ref{p}), the fitness becomes a
function of the Hamming distance $r$ only.  We note that the fitness $F$ is 
measured relative to that of a sequence with zero binding probability 
in any state $\alpha$. 

In a simple case, there are just two relevant cellular states. The 
{\em on} state favours expression of the gene, the {\em off} 
state disfavours it. It is then natural to assume selection coefficients 
of similar magnitude; here we take for simplicity 
$s = s^{\rm on} = -s^{\rm off} > 0$. 
We then obtain a {\em crater} landscape,
 \EQ
 F(r) = \frac{s}{1 + \exp[\ep (r - \rho^{\rm on})]}
      - \frac{s}{1 + \exp[\ep (r - \rho^{\rm off})]},
 \label{crater}
 \EE
with a high-fitness rim between $\rho^{\rm off}$ and
$\rho^{\rm on}$ flanked by  two sigmoid
thresholds; see fig.~1(a). The generic features of this fitness landscape are easy 
to interpret: the two-state selection assumed here favors intermediate 
binding strength (i.e., intermediate Hamming distances $r$) where binding
occurs and the gene is expressed in the {\em on state} but not in the {\em off}
state. Sequences with large Hamming distance $r > {\rho_{\rm on}}$ can bind
the factor neither in the {\em on} nor in the {\em off} state, while sequences
with $r < {\rho_{\rm off}}$ lead to binding in the {\em on} and the
{\em off} state. Both cases lead to misregulation of the downstream gene,
and hence to a lower fitness. We note that the key feature of these 
fitness landscapes, the sigmoid thresholds, is independent of the particular 
choices of $s^{\rm on}$ and $s^{\rm off}$. 

An even simpler fitness landscape is obtained if only the {\em on} state 
contributes significantly to selection, i.e., if 
$s = s^{\rm on} > 0$ and $s^{\rm off} = 0$. The crater landscape then 
reduces to the
{\em mesa} landscape discussed in~\cite{GerlandHwa:2002,Peliti:2002},
 \EQ
 F(r) = \frac{s}{1 + \exp[\ep (r-\rho^{\rm on})]},
 \label{mesa}
 \EE
which has a high-fitness plateau of radius
$\rho$ and one sigmoid threshold; see fig.~1(b). 
In this case, all sequences with sufficiently small Hamming distance to 
the target sequence ($r < \rho^{\rm on}$) have a high fitness.

In both cases, the parameters of the binding model have
a simple geometric interpretation:  $\ep$ gives  the slope
and the $\rho^\a$ give the positions of the sigmoid
thresholds in the fitness landscape. 
Eqs.~(\ref{crater}) and (\ref{mesa}) are again to be
understood as minimal models of 
fitness landscapes for binding sites, 
representing target sequence selection for 
a given level of binding ($\rho^{\rm off} < r < \rho^{\rm on}$)
and for sufficiently strong binding ($r < \rho^{\rm on}$), 
respectively. 
Despite its simplicity, this type of selection model 
based on biophysical binding affinities is nontrivial from
a population-genetic viewpoint since it leads to generic 
correlations between frequencies of nucleotides $a_i$ and $a_j$
within a site, see the Results section below. 
We will also study generalized models with correlations between two sites
generated by cooperative binding. On the other hand, 
these models neglect the context dependence
of the binding process through cofactors and chromatin structure.
However, they are a good starting point for order-of magnitude 
estimates of the adaptive evolution of binding sites. 

\section*{Mutation, selection, and genetic drift}

The rates of nucleotide point mutation show a great
variation, ranging from  $\mu \sim 10^{-4}$ per site and
generation for RNA viruses to values several orders of
magnitude lower in eukaryotes, e.g., $\mu \approx 2 \times
10^{-9}$ in {\em Drosophila}~\cite{Schlotterer.etal:1994}.
(Here we model mutation as a single-parameter Markov
process; we do not distinguish between transitions and
transversions.) The evolution of a sufficiently large 
population under mutation and selection can be described
in terms of the average fraction of  the population with a
given binding sequence. This so-called mean-field approach 
neglects the fluctuations due to finite population size
(genetic drift). It leads to the so-called {\em
quasispecies} theory~\cite{EigenMcCaskillSchuster:1989}.
For a population of sequences at a single binding site, 
the quasispecies
population equation can be written for the fraction $n(r,t)$ of
individuals
at Hamming distance $r$ from the target sequence at time $t$.
Along with a generalisation for two binding sites, it has
been analysed in detail in ref.~\cite{GerlandHwa:2002}. For the
mesa landscape, the stationary solution $n_\s (r)$ has been
found exactly \cite{Peliti:2002}. It depends only on the ratio
$s/\mu$ and describes a stable {\em polymorphic}
population, i.e., several sequence states coexist.  The
mean-field approach is valid as long as the stochastic
reproductive fluctuations are leveled out by mutations.
This requires absolute population numbers $N n_\s (r) \gg
1/\mu$ for all relevant $r$, a stringent condition on the
total population size $N$.

This paper is concerned with a different regime of population 
dynamics, as  described by the Kimura-Ohta
theory for finite populations evolving by stochastic
fluctuations (genetic drift) and
selection~\cite{KimuraOhta:1969,Kingman:1978,Ohta_Tachida:1990}. According to this theory,
a new mutant with a fitness difference $\Delta F$ relative to the
pre-existing allele could spread to fixation in the population. 
This is a stochastic process, whose rate constant is given by
\EQ
u = \mu N \, \frac{1 - \exp(- 2 \Delta F)}{1 - \exp(- 2 N \Delta F)}
\label{u}
\EE
in a diffusion approximation valid for $\Delta F \ll
1$~\cite{Kimura:1962}. Here $N$ is the {\em effective}
population size (with an additional factor 2 for diploid
populations). Eq.~(\ref{u}) has three well-known
regimes. For substantially {\em deleterious} mutations ($N
\Delta F \siml -1$), substitutions are exponentially
suppressed. {\em Nearly neutral} substitutions ($N |\Delta
F| \ll 1$)  occur at a rate $u \approx \mu$ approximately
equal to the rate of mutations in an individual. For
substantially {\em beneficial} mutations ($N \Delta F \simg 1$), the
substitution rate is enhanced, with $u \simeq
2 \mu N \Delta F$ for $N \Delta F \gg 1$.

In this picture, a population has a monomorphic majority
for most of the time and occasional coexistence of two
sequence states while a  substitution is going on. The time
of coexistence is $T \sim N$ for nearly neutral and $T \sim
1/ \Delta F$ for strongly beneficial substitutions.  The
picture is thus self-consistent for $T u \ll 1$, i.e., for
$\mu N \ll 1$. Asymptotically, it describes monomorphic
populations moving through sequence  space with hopping
rates $u$.

Introducing an {\em ensemble} of independent populations, this stochastic
evolution takes the form of a Master equation. For a single binding site,
we obtain
\begin{eqnarray}
\lefteqn{\mbox{$\frac{\partial}{\partial t}$} P (r,t) = }
\nonumber \\
& & (c-1)(\ell-r+1) \, u_{r-1,r} P(r-1,t) +
\nonumber \\
& &   (r+1) \, u_{r+1,r} P(r+1,t) - 
\nonumber \\
& &   [r \, u_{r,r-1} + (c-1)(\ell-r) \, u_{r,r+1}] P(r,t).
\label{master}
\end{eqnarray}
Here $P(r,t)$ denotes the probability of finding a
population at Hamming distance $r$ from the target
sequence, and $u_{r,r'}$ is given by (\ref{u}) with $\Delta
F = F(r') - F(r)$. The combinatorial coefficients arise
since a sequence at Hamming distance $r$ can mutate in
$(c-1)(\ell-r)$ different ways that increase $r$, and in $r$
ways that decrease $r$, where $c=4$ is the number of
different nucleotides. The stationary  distribution is
\EQ
P_\s (r) \sim \exp[ S(r) + 2 N F(r)].
\label{Pstat}
\EE
Here $S(r) = \log[({}^\ell_{\,r}) (c-1)^r / c^\ell]$ is the {\em
mutational entropy} (the log fraction of sequence states
with Hamming distance $r$)~\cite{Peliti:2002} and we have used the 
 exact result
$u_{r+1,r}/u_{r,r+1}=e^{2 (N-1) \Delta F}$. 
To derive (\ref{Pstat}), we then simply approximated 
$N-1$ by $N$. 
The form of $P_\s (r)$ reflects the selection
pressure, i.e., the scale $s$ of fitness differences in the
landscape $F(r)$. For neutral evolution ($2 s N = 0$),
the stationary distribution 
\begin{equation}
P^0_\s (r) \sim \sum \exp [S(r)]
\label{P0stat}
\end{equation}
is obtained from a flat
distribution over all sequence states. For moderate selection
($2 s N \sim 1$), $P_\s (r)$ results from a nontrivial
balance of stochasticity and selection. For strong
selection ($2 s N \gg 1$), $P_\s (r)$ takes appreciable
values only at points of near-maximal fitness, where $F(r)
\simg F_{\max} - 1/2 s N$. In this regime, the dynamics of
a population consists of beneficial mutations only, i.e.,
the system moves uphill on its fitness landscape.

The Master equation (\ref{master}) and the mean-field
quasi\-species equation  thus describe opposite asymptotic
regimes, $\mu N \ll 1$ and $\mu N \gg 1$, of the
evolutionary dynamics. Effective population sizes show a
large variation, from values of order $10^9$ in viral 
systems to $N\sim 10^6$ in {\em Drosophila} and
$N \sim 10^4 - 10^5$ in vertebrates. 
(These numbers bear some 
uncertainty; one reason is that $N$ varies across the
genome~\cite{BegunAquadro:1992}.) We conclude that the mean-field
quasispecies is well suited for viral systems, while
eukaryotes clearly show a stochastic dynamics of
substitutions.

\section*{Results and discussion}
\vspace{2mm}
\subsection*{Stationary distributions and nucleotide frequency correlations}

In the previous sections, we have expressed the fitness
landscape and the resulting population distributions as 
a function of the Hamming distance $r$ because it is a
convenient parameterization of the binding energy in the 
two-state model. In order to compare this approach to 
standard population genetics, it is useful to recast 
eq. (\ref{Pstat}) for the elementary sequence states 
$(a_1, \dots, a_l)$, 
\begin{equation}
P^0_\s(r) = 
  \sum_{(a_1,\dots,a_l) | r} {\cal P}_\s (a_1,\dots,a_l) \ ,
\end{equation}
where the sum runs over all sequence states at fixed
$r$. At neutrality, the distribution over sequence states
factorizes in the single nucleotide positions,
\begin{equation}
{\cal P}^0_\s (a_1,\dots,a_l) = \prod_{i=1}^{l}  \nu_0(a_i).
\end{equation}
In the specific case of the two-state model, 
$\nu_0(a_i)$ is simply a flat distribution over
nucleotides but it is obvious how this form can be
generalized to arbitrary nucleotide frequencies. 

According to eq. (\ref{Pstat}), the stationary distribution
under selection takes the form
\begin{equation}
\label{calPstat}
{\cal P}_\s (a_1,\dots,a_l) = 
  {\cal P}^0_\s (a_1,\dots,a_l) \exp[2 N F(r)]. 
\end{equation}
The salient point is that $F(r)$ is generically a strongly 
nonlinear function of $r$ due  to the sigmoid dependence
of the binding probability on $r$. An analogous statement
holds beyond the two-state approximation for the dependence
of $F$ on the binding energy $E$. Hence, even if
$ {\cal P}^0_\s (a_1,\dots,a_l)$ factorizes in the 
single nucleotide positions,  ${\cal P}_\s (a_1,\dots,a_l)$ 
does not. The selection introduces specific correlations
between the nucleotides: the fitness differences and, hence, the 
nucleotide frequencies at one position $i$ depend on all other 
$l-1$ positions in the motif.

\begin{figure*}[t!]
\includegraphics[width=.6\linewidth]{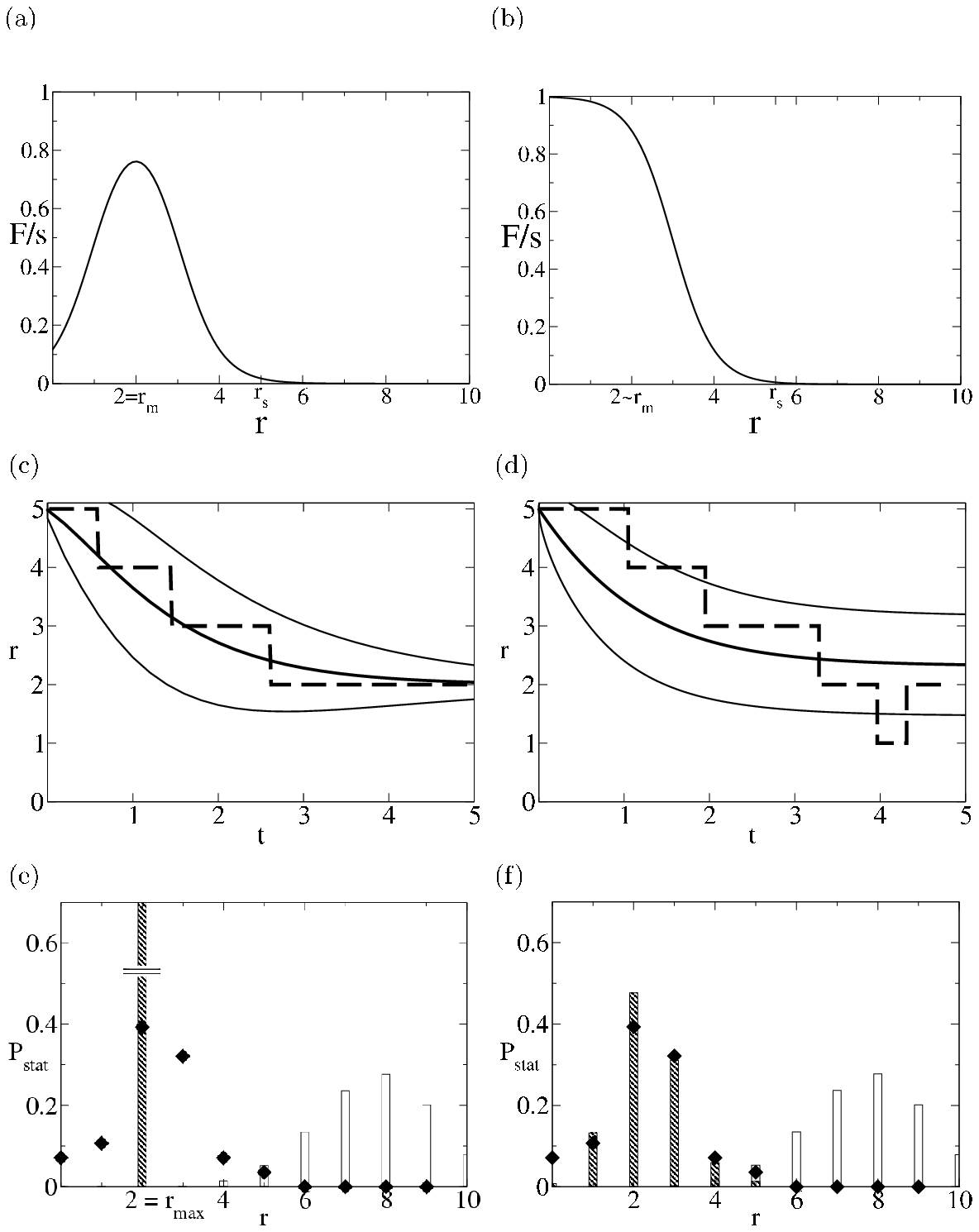}

\noindent
{\bf Figure 1:}
(a)~{\em Crater} landscape~(\ref{crater}) and
(b)~{\em Mesa} landscape~(\ref{mesa}), as a function
of the Hamming distance $r$ from the target sequence (within
the approximation of the two-state model). $r_{\max}$ gives the 
point where the binding probability reaches a maximum (crater landscape),
or else values close to $1$ (mesa landscape). $r_s$ approximately 
indicates the 
onset of selection, i.e. a binding probability appreciably different from 
zero.
(c)~Adaptive dynamics as a function of time $t$
measured in units of $1/(2 s \mu N)$ in the crater landscape
at strong selection ($s N = 100$). Single history $r(t)$ (dashed lines),
ensemble average $\bar r(t)$ (thick solid lines) and width
given by the standard deviation curves $\bar r(t) \pm \delta r(t)$ 
(thin solid lines).
(d) Same as (c) in the mesa landscape at moderate ($s N = 6.8$)
selection. 
(e)~Stationary ensembles $P_\s (r)$ of binding site
sequences with in the crater landscape at strong selection (filled bars) 
and for neutral evolution (empty bars). 
(f)~Same as (e) in the mesa landscape at moderate selection, together 
with the histogram of Hamming distances
of CRP site sequences in {\em E.~coli} from their consensus
sequence (diamonds, from~\cite{GerlandHwa:2002}).
\end{figure*}

\subsection*{Adaptive generation of a binding site}

We now apply the dynamics~(\ref{master}) to the problem of
adaptively generating a binding site in response 
to a newly arising selection pressure.  We study a case of
strong selection ($s N = 100$) in the crater fitness
landscape (\ref{crater}) with parameters $\ell = 10$, $\ep
= 2$, $\rho^{\rm on} = 3$, $\rho^{\rm off} = 1$ (implying
that the factor concentrations differ by a factor of 50),
and a case of moderate selection ($s N = 7$) in the mesa
landscape with parameters $\ell = 10$, $\ep = 1$, $\rho =
3.6$. (The mesa type may be most appropriate for factors
with multiple binding sites such as the CRP repressor in
{\em E.~coli}, where binding to an individual site is
negligible in the {\em off} state.) The fitness landscapes
for both cases are shown in fig.~1(a,b) in units of the 
selection pressure $s$. Substantially
beneficial mutations occur only on their sigmoid slopes,
i.e., in narrow ranges of $r$. The upper boundary of this
region is given by $r_s = \rho^{\rm on} + \log [s N ({\rm
e}^\ep - 1)]/ \ep$, which takes typical values $r_s = 5-7$.
In fig.~1(c,d), we show a sample history of adaptive
substitutions from $r=5$ to lower values of $r$, which are
close to the point $r_{\max}$ of maximal fitness. The
statistics of this adaptation is governed by the ensemble
$P(r,t)$; the average $\overline  r(t)$ and the standard
deviation $\delta r(t)$ appear also in fig.~1(c,d). In the
case of strong selection, the expected time of the adaptive
process is readily estimated in terms of the uphill rates
in (\ref{master}),
\EQ
\
T_s = \frac{1}{2 \mu N}
\sum_{r = r_{\max} + 1}^{r_s} \frac{1}{r (F(r-1) - F(r))},
\label{Ts}
\EE
and takes values of a few times $1/s \mu N$. We emphasize
again that this simple form depends only on the qualitative
form of the fitness landscape, namely, that weakly and
strongly binding sequence states are separated only 
by few point mutations. The conclusions are thus largely independent
of the details of the fitness landscape, which justifies using the two-state 
approximation. 

Can such a selective process actually happen? This depends on
the initial state of the promoter region in question {\em
before} the selection pressure for a new site sets in.
The region is approximated as an ensemble of
$L_1 = L - \ell + 1$ candidate sites undergoing {\em
independent} neutral evolution, i.e., the simultaneous 
updating of $\ell$ sites by one mutation is replaced by 
independent mutations. The length of the promoter region is denoted by $L$. 
At stationarity,  
the Hamming distance at a random site then
follows the distribution $P_
\s (r) \sim \exp[S(r)]$ shown as
empty bars in fig.~1(e,f). The minimal distance
$r_{\min}$ in the entire region 
is given by the distribution
${\cal P}(r) = Q_{\s}^{L_1} (r) - Q_{\s}^{L_1} (r+1)$,
where $Q_\s (r) = \sum_{r' \geq r} P_\s (r')$ is the 
cumulative distribution for a single site. ${\cal P}(r)$
is found to be strongly peaked, taking appreciable
values only in the range $\overline{r_{\min}}(\ell,L) \pm 1$
around its average. We assume selective evolution sets
in as soon as at least one site has a Hamming distance
$r \leq r_s$. This is likely to happen spontaneously
if $r_s \simg \overline{r_{\min}}(\ell,L)$, leading to 
a joint  condition on $\ell$, $L$, and $r_s$. 
For $r_s \siml \overline{r_{\min}}(\ell,L) - 1$,
there is a neutral waiting time before the onset of adaptation.
Its expectation value 
\EQ
T_0 = \frac{1}{\mu} \frac{Q_{\s}^{L_1 +1} (r_s + 1) }{
                    L_1 (r_s + 1) P_\s (r_s + 1)}
\label{T0}
\EE
is calculated in the appendix. It is generically
much larger than the adaptation time $T_s$, rendering
the effective generation of a new site less feasible.

The stationary distribution $P_\s (r)$ under selection is
given by (\ref{Pstat}) and shown as filled bars in
fig.~1(e,f). For strong selection, it is peaked at the
point $r_{\max}$ of maximal fitness. For
moderate selection, it
takes appreciable  values for $r=0 - 4$: the
binding site sequences  are {\em fuzzy}. Assuming that the 
CRP sites at different positions in the genome of {\em E.~coli} 
have to a certain extent evolved independently, we can fit 
$P_\s (r)$ with their  distance distribution 
(data taken from ~\cite{GerlandHwa:2002}).
At the values of
$\ep$ and $\rho^{\rm on}$ chosen, the two distributions
fit well, see fig.~1(f). This 
finding is discussed in more detail below.


\subsection*{Adaptation of binding cooperativity}

The cooperative binding of transcription factors involves
protein-protein interactions which may be specific to the
DNA substrate. These interactions often do not require
conformational changes of either protein involved and
depend only on few specific contact points. They result in
a modest energy gain of order  $3-4 k_B
T$~\cite{PtashneGann.book:2002}.  Hence, it is a
reasonable simplification to study the adaptive adjustment
of binding affinities using a simple generalisation of the
two-state binding model. We define the energies  $E_1 / k_B T = \ep
r_1$ and $E_2 / k_B T = \ep r_2$ for the binding of a single factor
and 
$E_{\rm pair} / k_B T = \ep[r_1 + r_2 - 2 (\gamma/\tilde \ell)
(\tilde \ell - \tilde r)]$ for the simultaneous binding of
both factors. The cooperativity gain is assumed to result
~from mutations at $\tilde \ell$ positions in the DNA
sequences of the factors, which encode the amino acids at
the protein-protein contact points. These mutations define
a Hamming distance $\tilde r = 0, \dots, \tilde \ell$ from
the target sequence for optimal protein-protein binding,
and $2 \gamma \ep/\tilde \ell$ is the binding energy per
nucleotide. Here we use the values $\ep = 2$, $\tilde \ell
= 6$ and $\gamma = 1$ but the qualitative patterns shown
below are rather robust. 

The resulting equilibrium probabilities  for the four
thermodynamic states $(--)$ (both factors
unbound), $(+-)$ and $(-+)$ (one factor bound), and $(++)$
(both factors bound) are
\EQ
\begin{array}{l}
q_{--},
\\
q_{+-}  =  q_{--} \exp[-\ep (r_1 -\rho_1)],
\\
q_{+-}  =  q_{--} \exp[-\ep (r_2 -\rho_2)],
\\
q_{++}  =  q_{--} 
     \exp[-\ep (r_1 + r_2 - \rho_1 - \rho_2 - 2 \gamma)],
\end{array}      
\label{q}
\EE
with the normalisation $q_{--} + q_{+-} + q_{-+} + q_{++} =
1$. The scaled chemical potentials $\rho_1$ and $\rho_2$
are independent  variables if the two sites bind to
different kinds of factors and are equal if they bind to
the same kind. As before, the binding probabilities
determine expression levels and, therefore,  the fitness.
Here we study only pairs of sites contributing additively
to the expression level in each cellular state, where we
have
\EQ
F = \sum_\a s^\a (q_{+-}^\a + q_{-+}^\a + 2 q_{++}^\a).
\label{F2}
\EE
Other important cases include activator-repressor site
pairs such as the famous {\em lac} operon~\cite{Muller-Hill.book:1996}, where
the transcription-factor induced expression level is
proportional to $q_{+-}$. The stochastic dynamics of
substitutions is straightforward to generalise; it leads to
a Master equation like (\ref{master}) for the joint
distribution $P(r_1,r_2,\tilde r, t)$. This
higher-dimensional equation can again be solved
exactly for its steady state 
\EQ
P_\s (r_1,r_2,\tilde
r) \sim \exp[S(r_1) + S(r_2) + S(\tilde r) + 2 N F(r_1,
r_2, \tilde r)].
\EE

\begin{figure*}[t!]
\includegraphics[width=.6\linewidth]{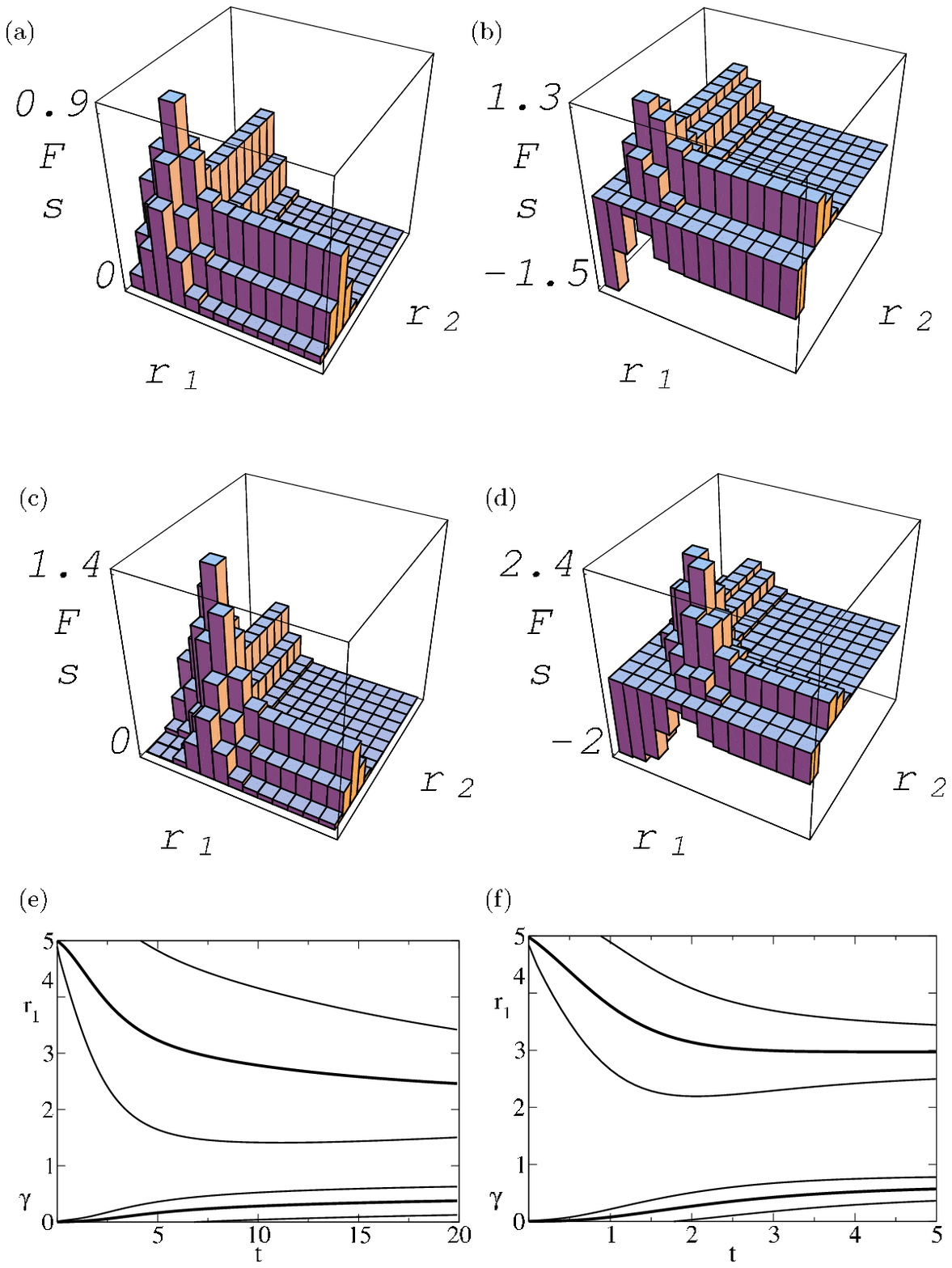}

\noindent
{\bf Figure 2:}
Genetic switch (left column), signal integration module
(right column).
(a,b)~Fitness landscape $F(r_1,r_2)$ without cooperativity
($\gamma = 0$).
(c,d)~Fitness landscape $F(r_1,r_2)$ with cooperativity
($\gamma = 1$). Next-nearest neighbour states $(r_1,r_2)$ and
$(r_1'= r_1 \pm 1,r_2' = r_2 \pm 1)$ of similar
fitness are linked by {\em compensatory} mutations if the
intermediate states $(r_1,r_2')$ and $(r_1',r_2)$ have lower
fitness.
(e,f)~Adaptive dynamics: ensemble averages 
$\overline r_1(t) = \overline r_2(t)$ and  
$\overline \gamma(t)$ (thick lines), ensemble
width given by 
$\overline r_{1} (t)  \pm \delta r_1(t)$ (same for $r_2$)
and $\overline \gamma (t) \pm \delta \gamma (t)$ 
(thin lines); cf.~fig.~1(e,f).
\end{figure*}

Here we discuss two simple examples of fitness landscapes
where binding cooperativity evolves by adaptation to
specific functional demands. A {\em genetic switch} with  a
sharp expression threshold is favoured in a system with a
single transcription factor having similar concentrations
in its  {\em on} and {\em off} cellular state.  As can  be
seen from eq.~(\ref{q}), cooperative binding can sharpen
the response of the binding probability to variations in
factor concentration,  $ q_{++} \sim 1/[1 + \exp(- 2 \ep
\rho + \dots)]$ versus $ p  \sim 1/[1 + \exp(-  \ep \rho +
\dots)]$ as given by (\ref{p}) for individual binding. 
Figs.~2(a,c) show the fitness landscape $F(r_1,r_2,\gamma)$
obtained from (\ref{q}) and (\ref{F2}) for $\rho^{\rm on} =
2.5$, $\rho^{\rm off} = 1.5$, and $s = s^{\rm on} = -s^{\rm
off}$. A simple {\em signal integration module} responds to
two different factors in four different cellular states,
({\em on, on}), ({\em on, off}), ({\em off, on}), ({\em
off, off}). Individually weak but cooperative binding leads
to expression of the  gene only if both factors are present
simultaneously. This case is favoured by a fitness function
of the form (\ref{F2}) with selection coefficients $s =
-s^{\rm off,off} = - s^{\rm on,off} = - s^{\rm off,on} =
s^{\rm on,on}/2$. The resulting fitness landscape
$F(r_1,r_2,\gamma)$ is shown in figs.~2(b,d) for chemical
potentials $\rho^{\rm on} = 3$, $\rho^{\rm off} = 1$ (for
each factor). 

In both cases, a pair of sites with weaker individual
binding ($r_1,r_2 = 3-4$) and cooperativity ($\gamma = 1$)
is seen to have a higher fitness than an optimal pair ($r_1
= r_2 = 2$) without cooperativity, as expected. Adaptive
pathways $\overline{r_{1,2}}(t)$ and $\overline \gamma (t)$
for strong selection ($s N = 100$) are shown in
fig.~2(e,f). Typical adaptation times $T_s$ are again a few
times $1/(s \mu N)$. A closer look reveals that this fast
adaptation sometimes leads to a {\em metastable} local fitness
maximum with some degree of cooperativity. {\em
Compensatory} mutations (see below) are then  required to
reach the global maximum, a process that may be
considerably slower. The fuzziness $\delta r_{1,2} (t)$ and
$\delta \gamma (t)$ observed in fig.~2(e,f) decays
on the larger time scale of compensatory mutations, 
reflecting  the presence of such metastable states.


\section*{Conclusions}
 
Transcription factors and their binding sites emerge as a
suitable starting point for quantitative studies of 
gene regulation.
Binding site sequences are short and their sequence space is
simple. Moreover, the link between sequence, binding affinity, 
and fitness is experimentally accessible. 
For a single site, the simplest examples are of
the {\em mesa}~\cite{GerlandHwa:2002} or of the {\em crater}
type, see fig.~1(a,b). Landscapes for a pair of sites with cooperative
binding interactions are of a similar kind as shown in
fig.~2(a-d). They can be used to predict the outcome
of specific single-site mutation experiments to a
certain extent. 

\subsubsection*{Fast adaptation may 
generate or eliminate a new binding site}

Despite this simplicity, the evolutionary dynamics of
binding sites is far from trivial, since it is governed, in
the generic case, by the interplay of three evolutionary
forces: selection, mutation, and genetic drift.  Here we
have focused on the dynamical regime appropriate for
eukaryotes, where the evolution can be approximated as a
stochastic process of substitutions. We find the possibility 
of selective pathways generating a new site in response to 
a newly arising selection pressure, starting from a neutrally
evolved initial state and progressing by point substitutions.
 Such a selective formation takes roughly $T_s \approx
\Delta r/(2 s \mu N)$ generations,  where $\Delta r$ is the
number of adaptive substitutions required. This number is
given by the Hamming distance between the onset of selection
and the point of optimal fitness, 
$\Delta r =  r_s - r_{\max}$, and takes values $2-3$ for typical
fitness landscapes; see fig. 1(a,b). For {\em
Drosophila melanogaster}, with  $\mu \approx 2 \times
10^{-9}$~\cite{Schlotterer.etal:1994}
and $N \approx 10^6$, the resulting 
$T_s$ is of the order of $10^6$
generations or $10^5$ years even for sites with a relatively
small selection coefficient  $s = 10^{-3}$. Such selective 
processes are faster than neutral evolution by a factor of about 
$1000$ and would allow for
independent generation of sites even after the split from
its closest relative {\em Drosophila simulans} about $2.5
\times 10^6$ years ago. Notice that new sites are more
readily generated in large populations. As discussed above,
generating  a new site may also require a neutral waiting
time $T_0$ until at least one candidate site in the
promoter region of the gene in question reaches  
the threshold distance $r_s$ 
from the target sequence, where
selection sets in. 
For site formation to be efficient, 
however, selection must be able to set in spontaneously, i.e., 
$T_0$ must not greatly exceed the adaptive time
$T_s$. This places a bound on the relevant length
$\ell$ of the binding motif that can readily form 
in a promoter region of length $L$.
Given $L \approx 300$, for example, a motif with $\ell = 8$
and $r_s = 3$ could still allow for  spontaneous
adaptive site formation. (For longer motifs, corresponding to groups 
of sites with fixed relative distance, this pathway 
would require promoter regions of much larger $L$.)
A more general case has recently been treated numerically in \cite{MacArthurBrookfield:2004},  
where the dependence of the neutral waiting time on the $G/C$ ratio of the 
initial sequence has been investigated. 
One may speculate that this adaptive dynamics is indeed one of the factors
influencing the length of regulatory modules in higher
eukaryotes.

 
Clearly, the present model also allows for pathways of
{\em negative selection} leading to the elimination of
spurious binding sites in regulatory or non-regulatory 
DNA where the binding has an adverse fitness effect. This 
is important since under neutral evolution, candidate sites
with a distance of at most $r_s$ from the target sequence
occur frequently on a genome-wide scale. A recent study has 
indeed found evidence for such negative selection from the
underrepresentation of binding site motifs over the entire 
genome~\cite{HahnStajichWray:2003}.

\subsubsection*{Binding sites under selection have nucleotide
frequency correlations}

We have shown that under stationary selection the frequencies
of nucleotides at any two positions of the binding sequence 
are correlated. For the two-state model, the correlations 
are the same for any pair of positions $i \neq j$ and can be computed
exactly from the joint distribution (\ref{calPstat}).
We emphasize that these correlations refer to an ensemble of independently
evolving (monomorphic) populations and are not to be confused with 
linkage disequilibria within one population. This finding limits 
the accuracy of bioinformatic weight
matrices, which are often assumed to factorize in the nucleotide
positions even in the presence of selection. 

\subsubsection*{Experimental tests: Binding site polymorphisms and phylogenies}

The predictions of our model lend themselves to a number of
experimental tests. In the dynamical regime appropriate for
eukaryotes ($\mu N \ll 1$), populations should be
monomorphic at most positions of their binding site
sequences and polymorphic at a few.  On the other hand, the
quasispecies model discussed in
refs.~\cite{GerlandHwa:2002,SenguptaDjordjevicShraiman:2002} (which assumes
$\mu N \gg 1$) may be most appropriate in viral systems. The
intermediate regime  $\mu N \sim 1$ with
frequent polymorphisms {\em and} genetic drift could be
realized in some bacterial systems and presents a challenge
for theory. Thus it would be very
interesting to compare the statistics of single-nucleotide
polymorphisms at binding sites in eukaryotes, bacteria, and
viruses. Polymorphism data are expected to contain evidence for
adaptive evolution. However, statistical tests of 
selection must be modified for promoter sequences
~\cite{JenkinsOrtoriBrookfield:1995,HahnStajichWray:2003}.
A recent study uses data on binding sites in three yeast species 
and deduces the rates of sequence evolution~\cite{Mosesetal:2003}. 

A complementary source of information are phylogenies of
binding sites.  Trees with functional
differences between branches contain information on the
generation of new sites or of interactions between sites
and on the time scales involved. In a tree for a conserved site
or group of sites with sufficiently long branches, 
the fuzziness of the sequences
observed on different branches is given by the 
ensemble $P_\s$ introduced above. For
strong selection, $P_\s$ lives on the {\em quasi-neutral}
network of  sequence states with maximal fitness, where two
neighbouring sequence states are  linked by  neutral
mutations or by pairs of {\em compensatory} mutations at
two different positions.  In the crater landscape
for a single site, this quasi-neutral network consists of
all sequences with  a fixed distance $r = r_{\rm max}$ from the
target sequence; see fig.~1(a). Beyond the two-state
approximation for binding energies, it will be smaller
since only some of the positions  are energetically
equivalent. For a group of sites, however, quasi-neutral
networks can be larger since compensatory mutations can
also take place at positions on different sites as shown in
fig.~2(d) for the example of a signal integration module.
This is consistent with  experimental evidence that the
sequence divergence between {\em Drosophila melanogaster}
and  {\em Drosophila pseudoobscura} involves compensatory
mutations and stabilising  selection between different
binding sites~\cite{Ludwig.etal:2000}.  


For weaker selection, site fuzziness increases further
since $P_\s$ extends beyond the sequence states of maximal
fitness and is influenced by mutational entropy. As shown
in fig.~1(f), one can explain in this way the observed fuzziness
in CRP sites of {\em E.~coli}. It would then
reflect different evolutionary histories of independent
populations, rather than sampling in one polymorphic
population as in the quasispecies picture of
refs.~\cite{GerlandHwa:2002,SenguptaDjordjevicShraiman:2002}. 
(In a mean-field 
quasispecies, appreciable fuzziness occurs only
for selection coefficients $s \sim \mu$, minute in
other than viral systems.)
However, the data are also compatible with
strong selection if the selection coefficients $s^\alpha$, 
and hence the value of $r_{\max}$,
vary between different genes. Clearly, comparing 
$P_\s$ with the distribution of sites in a single genome
requires the assumption that the evolutionary histories
of sites at different positions are at least to some extent
independent. Future data of orthologous sites in a sufficient
number of species will be more informative. Thus, further
experimental evidence is needed to clarify the role of
mutational entropy in the observed fuzziness.

\subsubsection*{Evolvability of binding sites}

The present work was aimed at obtaining some insight into
the molecular  mechanisms and constraints underlying the
dynamics of complex regulatory  networks, thereby
quantifying the notion of their {\em evolvability}.
The programming of binding sites and of  cooperative
interactions between them is found to provide efficient modes of
adaptive evolution whose tempo can be quantified for the 
case of point mutations. The formation of complicated
signal integration patterns and of multi-factor interactions in higher eukaryotes, 
however, requires generalizing our arguments in
two ways. There are further modes of sequence evolution 
such as slippage events, insertions and deletions, large 
scale relocation of promoter regions, and recombination. 
Our ongoing work is aimed at quantifying their relative importance
in terms of substitution rates. Moreover, there are also more general
fitness landscapes describing, e.g., binding sites
interacting via the expression level of the regulated gene
(such as activator-repressor site pairs) and the coupled
evolution of binding sites in different genes. 

The rapid evolution of networks hinges upon the existence of
adaptive  pathways for these formative steps with a characteristic
time scale  $T_s \sim 1/ (s \mu N)$  much smaller than $T_0
\sim 1/\mu$, the time scale of neutral  evolution. The
presence of these two time scales has a further interesting
consequence. If the selection pressure on an existing site
ceases, that site will disappear on the larger time scale
$T_0$. It is possible, therefore, that large existing
networks have accumulated a considerable number of {\em
redundant} regulatory interactions acquired by selection in
their past. This may be one factor contributing to their
robustness against perturbations.

\section*{Methods - Neutral evolution of binding sites}

To estimate the average neutral waiting time $T_0$, we
study the mutation dynamics in the restricted range 
$r = r_s+1, ..., \ell$, allowing mutations from $r_s+1$
to $r_s$ but suppressing mutations from $r_s$ back to
$r_s+1$. We evaluate the time-dependent solution 
$P(r,t)$ of the Master equation (\ref{master}) with 
the initial condition $P(r,0) = P_\s(r)$, 
and the resulting cumulative probability 
$Q(t) = \sum_{r \geq r_s +1} P(r,t)$. 
The current across the lower boundary, 
$J(t) = \mu (r_s+1) P(r_s+1,t) = -{\rm d}Q/{\rm d}t$, determines
the waiting time for a single site,  
\EQ
T_0 = \int_0^\infty \,{\rm d}t\, t \, J(t) 
= \int_0^\infty \,{\rm d}t\, Q(t).
\label{Tint}
\EE
This is formally solved by expanding in 
eigenfunctions of the mutation operator.
In the case relevant here, the system remains close
to equilibrium since the boundary current is much smaller 
than typical currents for $r \geq r_s$.
Hence,  $P(r,t) \approx P_\s(r) \exp(- \lambda t)$ with 
$\lambda = J(0)/Q(0) 
 = \mu (r_s +1) P_\s (r_s + 1) /  Q_\s (r_s+1)$. 
We conclude that the  waiting time for a single site 
is positive with probability $Q_\s (r_s + 1)$, following
a distribution $\sim \exp(- \lambda t)$, and 0 otherwise. 
The resulting expectation value is 
$T_0 = Q_\s (r_s + 1)/\lambda$.
For $L_1$ independent sites, the distribution of positive 
waiting times is still exponential, and $T_0$ is given
by an expression of the form (\ref{Tint}) with a total 
boundary current 
$J(t,L_1) = {\rm d} Q^{L_1} (t) / {\rm d} t$. This
yields 
$T_0 = Q_{\s}^{L_1}(r_s+1)/ L_1 \lambda$ as given by
(\ref{T0}).  The average 
waiting time (in units of $1/\mu$) becomes large for 
values of $r_s$ in the 
tail of the distribution ${\cal P} (r)$, where 
$Q^{L_1}_\s (r_s+1) \approx 1$. This is the case
for $r_s \siml \overline{r_{\min}}(\ell,L) -1$.

\section*{Authors contributions}

JB carried out analytical and numerical work, SW 
performed numerical work and data processing. 
ML conceived of the study, carried out analytical work, 
and coordinated the project. All authors read and approved
the final manuscript.

\section*{Acknowledgements}
  \ifthenelse{\boolean{publ}}{\small}{}
We are indebted to  N. Rajewsky and D. Tautz for
interesting discussions, and to P. Arndt 
and U. Gerland for a critical reading of the manuscript. 
This work has been supported by DFG grant LA 1337/1-1.


{\ifthenelse{\boolean{publ}}{\footnotesize}{\small}
 \bibliographystyle{bmc_article}  
  \bibliography{../../text/regulatory} }     


\ifthenelse{\boolean{publ}}{\end{multicols}}{}


\end{bmcformat}
\end{document}